\documentclass[useAMS,usenatbib]{mn2e}
\usepackage{epsfig}
\usepackage{graphicx}

\def\gtrsim{\lower 2pt \hbox{$\, \buildrel {\scriptstyle >}\over
{\scriptstyle \sim}\,$}}
\def\lesssim{\lower 2pt \hbox{$\, \buildrel {\scriptstyle <}\over
{\scriptstyle \sim}\,$}}

\DeclareGraphicsExtensions{.gif,.jpg,.pdf,.png}

\title[Galactic Coronae in the ICM]{Galactic Coronae in the Intracluster Environment: Semi-confined
Stellar-feedback-driven Outflows}
\author[]{
Zhankui Lu$^{1}$ \thanks{E-mail:lv@astro.umass.edu}
and Q. Daniel Wang$^{1}$\\
$^{1}$ Department of Astronomy at University of Massachusetts, Amherst, 01003, USA}

\begin{document}
\date{}
\maketitle

\begin{abstract}
Recently X-ray observations have shown the common presence of compact galactic 
coronae around intermediate-mass spheroid galaxies embedded in the 
intracluster/intragroup medium (ICM). 
We conduct 2-D hydrodynamic simulations to study the quasi-steady-state 
properties of such coronae as the natural products of the ongoing distributed 
stellar feedback semi-confined by the thermal and ram pressures of the ICM. 
We find that the temperature of a simulated corona depends 
primarily on the specific energy of the feedback, consistent with the lack of
the correlation between the observed hot gas temperature and K-band 
luminosity of galaxies. The simulated coronae typically represent 
subsonic outflows, chiefly because of the semi-confinement. As a result, 
the hot gas density 
increases with the ICM thermal pressure. The ram pressure, on the other 
hand, chiefly affects the size and lopsidedness of the coronae. The density
increase could lead to the compression of cool gas clouds, if present,
and hence the formation of stars. The increase also enhances radiative
cooling of the hot gas, which may fuel central supermassive black holes,
explaining the higher frequency of active galactic nuclei observed in clusters than in 
the field. 
The radiation enhancement is consistent with a substantially 
higher surface brightness of the X-ray emission detected from 
coronae in cluster environment. The total X-ray luminosity of a
corona, however, depends on the relative importance of the surrounding 
thermal and ram pressures. These environment dependences should at least partly explain the large 
dispersion in the observed diffuse X-ray luminosities of spheroids
with similar stellar properties. 
Furthermore, we show that an outflow powered by the distributed 
feedback can naturally produce a positive radial gradient 
in the hot gas entropy, mimicking a cooling flow.
\end{abstract}

\section{Introduction}

Though consisting of primarily old stars, galactic spheroids (bulges of Sb-Sa
spirals as well as S0 and elliptical galaxies) are a major
source of stellar feedback in form of mass loss and Type Ia supernovae (e.g., 
Ciotti et al. 1991; Knapp et al. 1992;  Mannucci et al. 2004). The specific
energy of this feedback predicts that it should present primarily in X-ray-emitting 
hot gas. Indeed, such hot gas has been detected in and around spheroids,
which typically contain little cool gas. However, it has been shown repeatedly 
that the X-ray-inferred hot gas mass and energy are far less than 
the empirical predictions from the feedback inputs (e.g., 
\citealt{David06,Li06,Li07a,Li07b,Wang10}).   
This missing stellar feedback has most likely escaped into large-scale 
galactic halos, where the gas becomes too tenuous to be detected in existing 
X-ray imaging observations (e.g.,
\citealt{Tang09a} and references therein).
The implication of this scenario is profound, because the injection of the
mass and energy into the halos could strongly affect the ecosystem of the 
galaxies and hence their evolution \citep{Tang09a}.

While the above qualitative picture seems clear, there are key issues that still need to be
addressed to understand both the stellar feedback itself and its interplay with
the environment. Significant uncertainties are still present in the mass and energy input 
rates of stars (\S~2). 
The energy input rate is inferred from observations of SNe in a large sample of galaxies of diverse
optical and near-IR luminosities as well as types (e.g., \citealt{Mannucci et al. 2005}), 
assuming a certain explosion mechanical energy deposited into the interstellar space. The stellar mass loss
rate is based on the modeling of the $12{\rm~\mu m}$ emission from the circumstellar medium of evolved stars
(e.g., \citealt{Knapp et al. 1992}). These semi-empirical rates, uncertain by a 
factor of at least $\sim 2$ for individual galaxies, can in principle be 
directly constrained by the measured temperature and luminosity of 
galactic coronae. Indeed, detailed simulations have been conducted for relatively
isolated ``field'' spheroids and have been compared with observations \citep{Tang09a,Tang09b,TangWang10}, which have led to a qualitative understanding of the feedback
processes and effects on X-ray measurements. 
In particular, the feedback model expects that the specific energy should not 
change substantially from one spheroid to another, which is consistent with 
little correlation between the measured temperatures and 
K-band luminosities $L_K$ (e.g., \citealt{David06, Sun et al. 2007}). 
But the measured temperatures ($\lesssim$ 1 keV)
are substantially lower than the expected value from simulations. The
measured iron abundances of coronal gas are also typically lower than 
expected for the Ia SN-enriched gas (\S~2).
At least part of these discrepancies can be accounted for by various 
3-D effects of the Ia SN heating, which produces very low-density, hot, 
fast-moving, and enriched bubbles that hardly radiate. When the gas in 
these bubbles finally mixed with the material from the stellar mass loss 
at large radii, the X-ray emission becomes 
too weak and diffuse to be effectively detected 
\citep{Tang09a, Tang09b,TangWang10}. The observed X-ray 
emission thus gives only a biased view of the coronae. 
We expect that this bias should be minimal for a corona 
in the ICM, however. The high external thermal and ram pressures
tightly confine such coronae, resulting in a low outflow speed and hence 
relatively local mixing of the feedback materials. 
Furthermore, one can better characterize the ICM environment 
from observations, important for a self-consistent modeling of a corona. 
Therefore, coronae embedded in the ICM are ideal sites to better constrain 
the feedback and its interplay with the environment.

There have also been significant efforts in studying stellar feedback-powered coronae embedded in the ICM, 
mostly focusing on the ram-pressure stripping of hot gas
(e.g., \citealt{Acreman et al. 2003, Stevens et al. 1999, 
Toniazzo and Schindler 2001}). It is shown that the global morphological and 
integrated properties, such as gas mass and luminosity, are strongly influenced 
by the environment. A set of 2D simulations done by \citet{Stevens et al. 1999}
show that galactic coronae can be maintained by stellar feedback in poor clusters 
while be efficiently stripped in rich ones. 
\citet{Acreman et al. 2003} and \citet{Toniazzo and Schindler 2001} simulated galaxies 
falling into clusters and demonstrated that a galactic corona reached 
a cyclic "stripping replenishment" dynamics due to the periodic orbital motion of
the host galaxy as well as the competing processes such as stripping and stellar feedback.   

We focus on modeling coronae in and around intermediate-mass spheroids that 
are embedded in the ICM. Such a system is relatively simple with minimum effects 
due to the feedback from AGNs and to the 
radiative cooling of hot gas. We expect that the hot gas is in a
quasi-steady, subsonic outflow semi-confined by the thermal and ram-pressures of 
the ICM. This state should be only sensitive
to the local properties of the ICM  (see \S~2 for more discussion),
avoiding large uncertainties in modeling the history of the galactic feedback and
the evolution of the environment, as would be needed for a field spheroid
\citep{Tang09a}. The simulations can also be 
compared with an increasing number of X-ray detections of 
galactic coronae of such spheroids (e.g., \citealt{Sun et al. 2007}), 
leading to an improved understanding of the feedback itself and its 
interplay with the environment. In particular, we examine 
the dependence of the corona properties on the specific energy of the 
stellar feedback and on the thermal and ram pressures of the ICM and
check how measurements (e.g., temperature, surface brightness, overall 
luminosity and morphology) may be made to infer the parameters that cannot 
directly observed (e.g., the ICM thermal and 
ram pressures local to an individual galaxy). Here, we will present 2-D 
simulations only, which allow for an efficient exploration of 
a large parameter space.
The paper is organized as follows: We briefly describe our numerical model 
and setup in \S~2 and present results in \S~3; We discuss their
implications in \S~4; Finally in \S~5, we summarize our conclusions. 

\section{Model and Simulation Setup} 
\label{s:setup}
\subsection{Model Galaxies}
\label{ss:setup_galaxy}
Our model galaxy is composed of a stellar spheroid component and a dark 
matter halo. We use the spherical Hernquist density profile 
\citep{Hernquist 1990} to represent the stellar mass distribution:

\begin{equation} 
 \rho_s(r) = {M_s \over 2\pi a^3} {a^4 \over r(r+a)^3},
\end{equation}
where $M_s$ is the total stellar mass, and $a$ is the scale radius. This density profile results in a gravitational potential
\begin{equation}
\phi(r) = -{GM \over (r+a)}.
\end{equation}
The above stellar mass distribution resembles the de Vaucouleur's Law; 
The relation between the half-light radius $R_e$ and the scale radius is 
$R_e = 1.8513a$.

We characterize the dark matter halo with the NFW profile \citep{Navarro Frenk and White 1997}, 
\begin{equation}
\rho_d(r) = {\rho_0 \over (r/r_d)(1+r/r_d)^2},
\end{equation}
where $r_d$ is the scale radius of the dark halo, and $\rho_0$ is defined as
\begin{equation}
\rho_{0} = {1\over3} \rho_{crit}\Omega_m\Delta_{vir} {c^3 \over \ln(1+c)-{c\over (1+c)}},
\end{equation}
in which $\rho_{crit}$ is the critical density of the universe. The dark halo 
has a mass $M_{vir}$ within the virial radius $r_{vir}$, which is defined to 
have a density that is $\Delta_{vir}$ times the mean matter density of the universe $\rho_{crit}\Omega_{m}$. 
We adopt $\Delta_{vir}$ to be $250$. Therefore, we have the relation 
\begin{equation}
M_{vir} = {4\pi \over 3} r_{vir}^3 \Delta_{vir} \rho_{crit} \Omega_{m},
\end{equation}
where $c$ is the concentration factor defined as $c = { r_{vir} \over r_d }$ and is related to $M_{vir}$. 
We set $c=13$ \citep{Eke et al. 2001}. 
Thus, for a given cosmology and a given $M_{vir}$, the mass profile of the dark halo is totally determined.

\subsection{Stellar Feedback}
\label{ss:setup_feedback}
The stellar mass and energy feedback in spheroids are dominated by the 
mass loss and Ia SNe of evolved stars, respectively. We neglect the 
energy input from the random motion of stars and hence their ejecta. 
The total energy released by type Ia SNe is
\begin{equation}
\dot{E} = E_{SN} n_{SN} \Bigl( {L_K \over 10^{10}L_{K,\odot}} \Bigr),
\end{equation}
where $n_{SN} = 0.00035 {\rm~yr^{-1}}$ for E/S0 galaxies according to \citet{Mannucci et al. 2005}. 
The empirical mass input rate from the stellar mass loss is
\begin{equation}
\dot{M} = m \Bigl( {L_K \over 10^{10}L_{K,\odot}} \Bigr),
\end{equation}
where $m = 0.021{\rm~M_{\odot}yr^{-1}}$ according to \citet{Knapp et al. 1992}.  
Assuming that the mechanical energy of each SN is $10^{51}{\rm~erg}$, 
the specific energy of stellar feedback is $\beta = {\dot{E}\over \dot{M}} \sim 5{\rm~keV}$ per particle. 
To account for the uncertainties in these rates and assumptions, we also sample three different lower values
of the specfic energy for comparison with observations (Table 1).
We assume that the energy and mass inputs follow the distribution of the stellar mass.

In addition, each Ia SN produces $\sim 0.7M_{\odot}$ of iron ejecta. We assume that the iron abundance of 
the mass loss from stars is solar. If the ejecta is fully and instantaneously 
mixed with the mass loss, the expected iron abundance relative to 
the solar value is then 
\begin{equation}
\Bigl( {n_{sn}M_{Fe}\over m} \Bigr) / Z_{\odot} \sim \Bigl( {0.0018 \times 0.7 M_{\odot} \over 0.2 M_{\odot}} \Bigr) /Z_{\odot} = 5.5. 
\end{equation}
However, in observations, supersolar metallicity is quite rare. \citet{Tang09b} and \citet{TangWang10} have shown that 
Ia SN ejecta may not be efficiently mixed with stellar mass loss on microscopic scale, 
resulting in a low effective metallicity of the ISM. While our focus is on the environmental effect.
In our simulation, we set the iron abundance of the input mass to the
solar value.

\subsection{Simulation Setup}
\label{ss:setup_setup}
The gas hydrodynamics and metal abundance distribution with both the 
stellar feedback and radiative cooling can be described by the following 
equations:
\begin{eqnarray}
{\partial \rho_g \over \partial t} + \bf \nabla \cdot (\it \rho_g \bf v) \it &=& S_m                                   \nonumber  \\
{\partial (\rho_g \bf v) \over \it \partial t} + \bf \nabla \cdot (\it \rho_g \bf vv \it + P[I]) &=& \rho_g \bf g \it  \nonumber  \\
{\partial (\rho_g e) \over \partial t} + \bf \nabla \cdot \it [(\rho_g e + P)\bf v \it] &=& S_e - n_e n_i \Lambda(T,Z) + \rho_g \bf g \cdot v \it \nonumber  \\
{\partial (\rho_g X_{iron}) \over \partial t} + \bf \nabla \cdot \it (\rho_g X_{iron} \bf v \it ) &=& S_{iron} .
\end{eqnarray} 
The first equation is mass conservation law, with $\rho_g$ denoting the mass density of the coronal gas.
The second is momentum equation. $P$ is gas pressure and $\bf g$ is gravitational acceleration.
The third one is energy equation. $e$ stands for the specific energy of the gas, including both thermal 
and kinetic components. The second term to the right is cooling rate. 
We adopt the cooling curve from \citet{Sutherland and Dopita 1993}, 
assuming an optically-thin thermal plasma in collisional ionization equilibrium. 
For calculation of the radiation in a specific band, we use the Mekal model, extracted from the X-ray spectral 
analysis software XSPEC.
We use the fourth equation to keep track on the iron mass fraction, which is denoted by $X_{iron}$. 

We conduct our simulations with the FLASH code \citep{Fryxell et al. 2000},
an Eulerian astrophysical hydrodynamics code with the adaptive mesh 
refinement (AMR) capability. The simulated region is fixed in the galaxy-rest
frame using cylindrical coordinates, with $z$ ranging from $-50{\rm~kpc}$ to $50{\rm~kpc}$ and
the radius from $0$ to $50{\rm~kpc}$. The axis of the cylinder is 
through the center of a simulated spheroid and along the direction
of its motion. The upper and lower boundary conditions are 
fixed so that the ICM flows in and out the simulation region at a constant speed. This
mimics the motion of galaxies through a local cluster environment.
We apply reflection boundary condition at $r=0$ and 
diode boundary condition, which only allows gas to flow out, to the right side of the simulation region.
Compared with the simulation box, which is $50{\rm~kpc}$ by $100{\rm~kpc}$, the coronae are only on the
order of $1{\rm~kpc}$ to $10{\rm~kpc}$ across. 
Data near the outer regions will be excluded in our analysis to
avoid any potential artifacts introduced by the assumed outer
boundary condition of the simulations. 
Also, the outer region will not be shown in the following images. 
We allow the resolution to reach $0.1{\rm~kpc}$, so that 
the small coronae can be well resolved. 

When a simulation starts, there is no interstellar gas in the galaxy. 
As the simulation progresses,
the stellar feedback gradually accumulates in and around the spheroid to form
a corona, which is characterized by its higher iron abundance. In the mean time,
the ram-pressure and turbulent motion strips gas at the outer 
boundaries of the corona. We end the simulation when it reaches a 
statistically quasi-steady state. Empirically, the time to reach
such a state is $\tau_g\sim0.2$ Gyr, 
while the time for the ICM to pass the simulation region ranges from $0.05$ Gyr to $0.2$ Gyr. 
Representative results are all extracted 
from the simulations after this time.

The presence of a local quasi-steady state is a reasonable assumption
for a compact corona. As a galaxy moves through a cluster,
the ICM condition can of course change drastically. But the time scale for 
such a change is typically much longer than the dynamic time for the 
corona to adapt the local environment. For a cluster of 
a characteristic size of $\sim 1$ Mpc and temperature of 2 keV, for example, 
the crossing 
time for a galaxy moving roughly at the sound speed is $\tau_{c1} \sim 2$ Gyr. 
In contrast, for a corona of a typical size $\sim 5$ kpc and temperature 
$\sim 0.8$ keV, the sound crossing time is only $\tau_{c2} \sim 20$ Myr. 
Even if a corona is totally destroyed at some point(e.g. at the central region of a cluster), 
the re-building time scale $\tau_g$, as dicussed above, is still shorter than 
the environment change time scale. Therefore, the local quasi-steady state
is a reasonable assumption for characterizating the
environmental impact on galactic coronae. 


\section{Results}
\label{s:results}
We have simulated a set of cases to characterize the dependence on key 
parameters. Table 1 lists our adopted model parameter values.
The different combinations of the ICM density ($n_{i}$, the number density of all particles)
and temperature ($T_{j}$) 
as well as the Mach number ($\mathcal M_{k}$) and specific 
energy ($\beta_{l}$) of the model galaxy form a set of 48 cases.

\begin{table*}
\label{t:model_p}
 \centering
 \begin{minipage}[c]{0.5\linewidth}
  \centering
  \caption{Model Parameters}
  \begin{tabular}{@{}lcccc@{}}
 \hline
 \hline
 Model galaxy
                   & Stellar Mass($10^{11}{\rm~M_{\odot}}$)    & $2.0$ \\
                   & Dark Halo Mass ($10^{11}{\rm~M_{\odot}}$) & $40$  \\ 
 \hline
 ICM properties
                   & ICM density ($10^{-4} {\rm~cm^{-3}}$) & $3.3(n_1)$, $10(n_2)$  \\
                   & ICM temperature ($10^{7} {\rm~K}$)    & $2.0(T_1)$, $6.0(T_2)$ \\
                   & Iron abundance ($\rm~Z_{\odot}$)      & $0.3$                  \\
                   & Mach number                           & $0.6(\mathcal M_{1})$, $1.2(\mathcal M_{2})$, $1.8(\mathcal M_{3})$ \\
 \hline
 Stellar feedback
                   & Mass loss rate  (${\rm~M_{\odot}} / 10^{11}{\rm~M_{\odot} /yr}$) 
                                                           & $0.32$                 \\
                   & Specific energy (keV)                 & $1.2(\beta_1)$, $1.8(\beta_2)$, $3.0(\beta_3)$, $4.8(\beta_4)$ \\
                   & Iron abundance ($\rm~Z_{\odot}$)      & $1.0$                  \\
 \hline
 \hline 
\end{tabular}
\end{minipage}
\end{table*}

Here we present the gas properties extracted from the
simulations. We first detail the results for a representative 
case $n_1T_2 \mathcal M_2 \beta_2 $ (\S~\ref{ss:model_r}) and then 
discuss  the similarity and significant variation among the different cases (\S~\ref{ss:model_v}).

\subsection{Representative Case}
\label{ss:model_r}
\begin{figure}
   \centering
   \begin{minipage}[c]{0.48\linewidth}
      \centering
      \includegraphics[width=\linewidth]{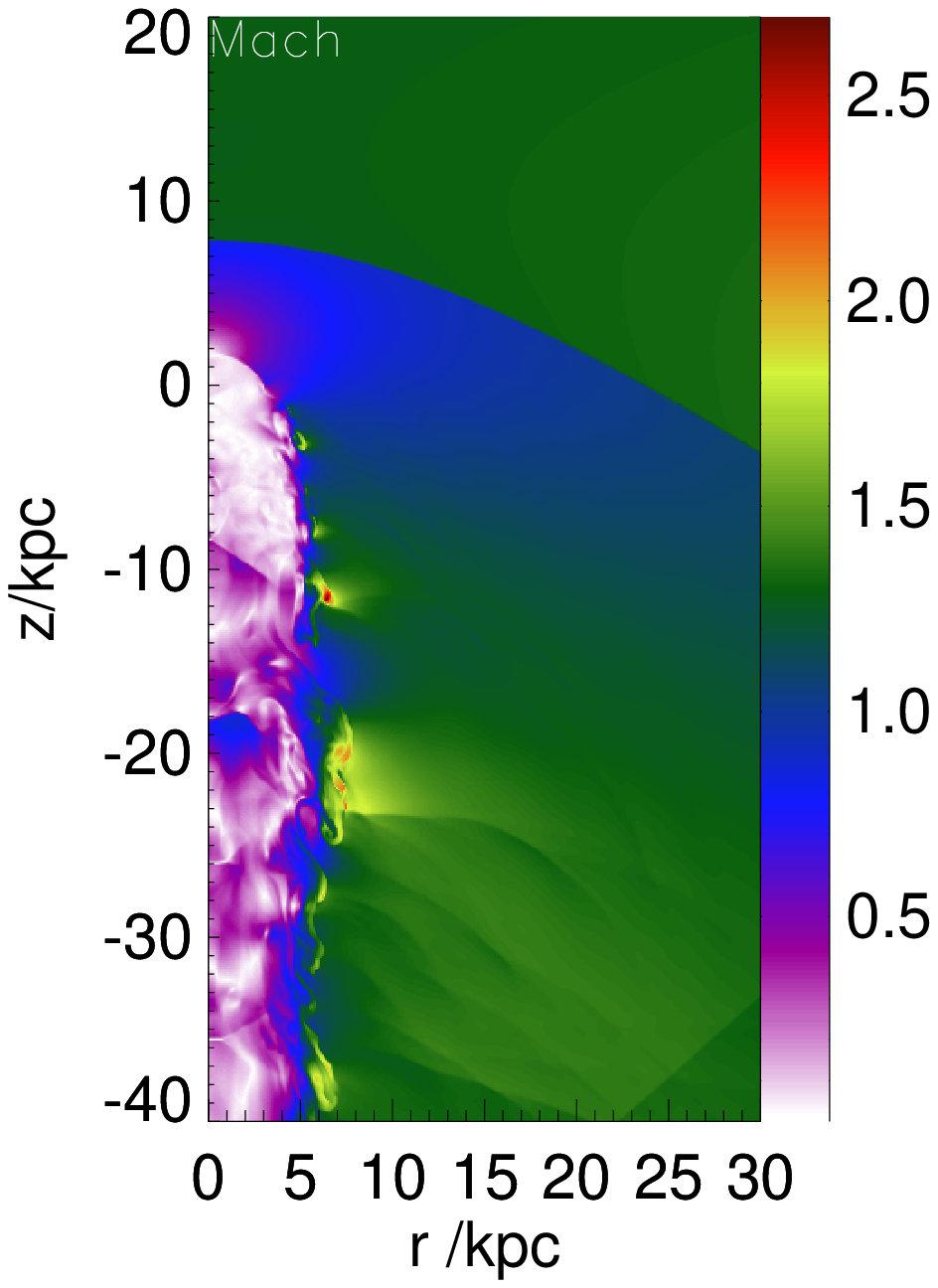}
   \end{minipage}%
   \hspace{0cm}%
   \begin{minipage}[c]{0.48\linewidth}
      \centering
      \includegraphics[width=\linewidth]{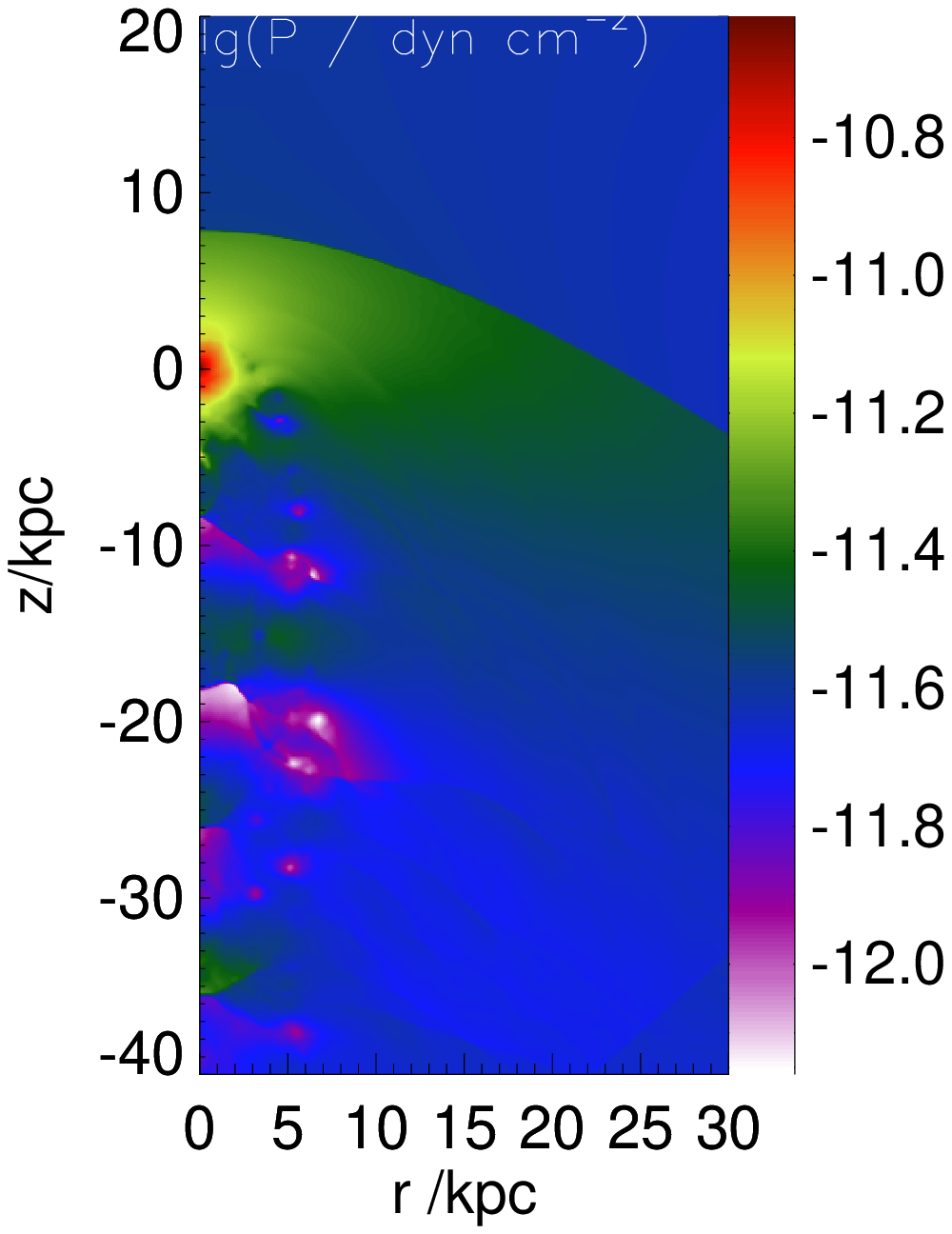}
   \end{minipage}\\[0pt]
   \begin{minipage}[c]{0.48\linewidth}
      \centering
      \includegraphics[width=\linewidth]{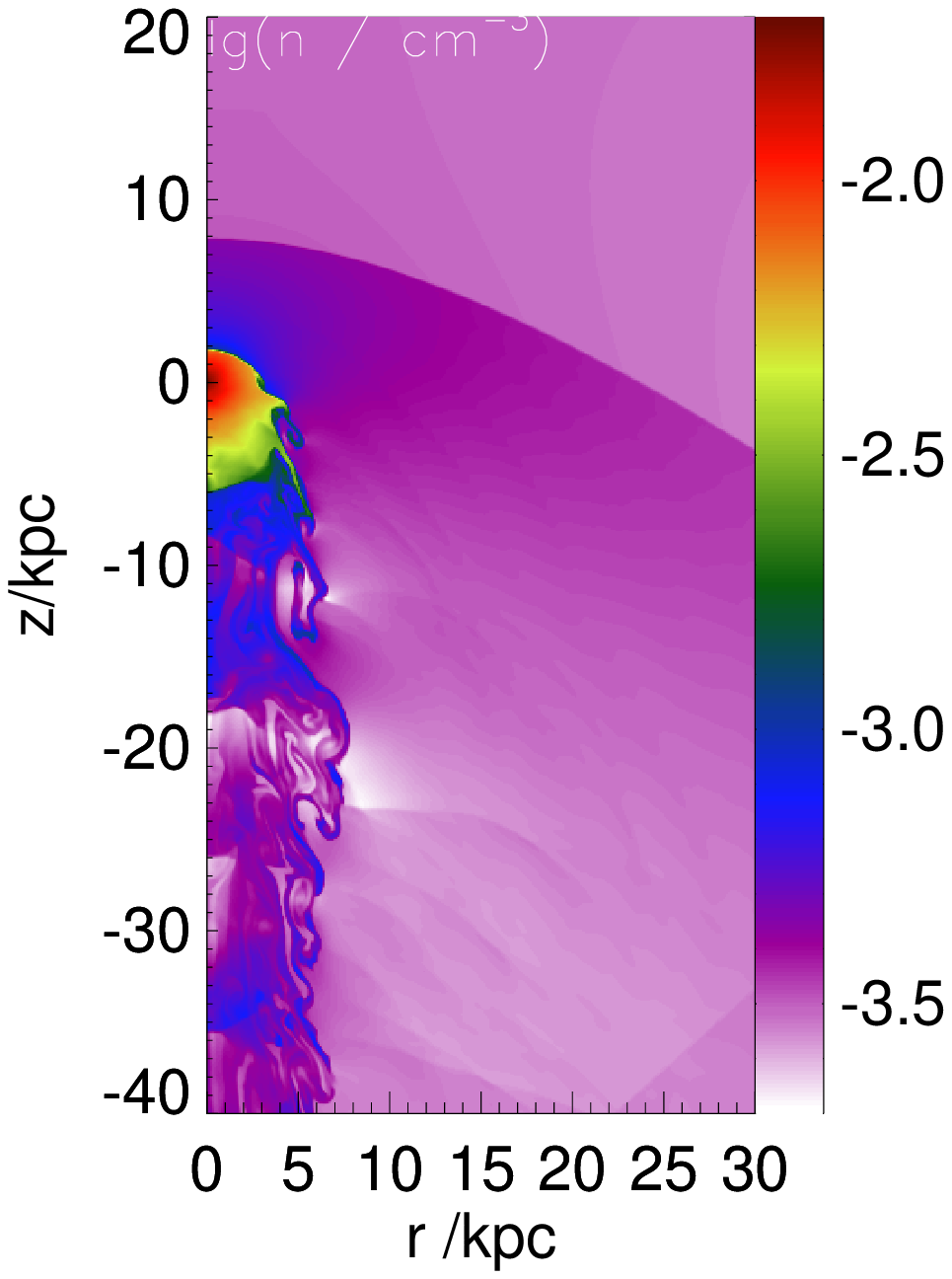}
   \end{minipage}%
   \hspace{0cm}%
   \begin{minipage}[c]{0.48\linewidth}
      \centering
      \includegraphics[width=\linewidth]{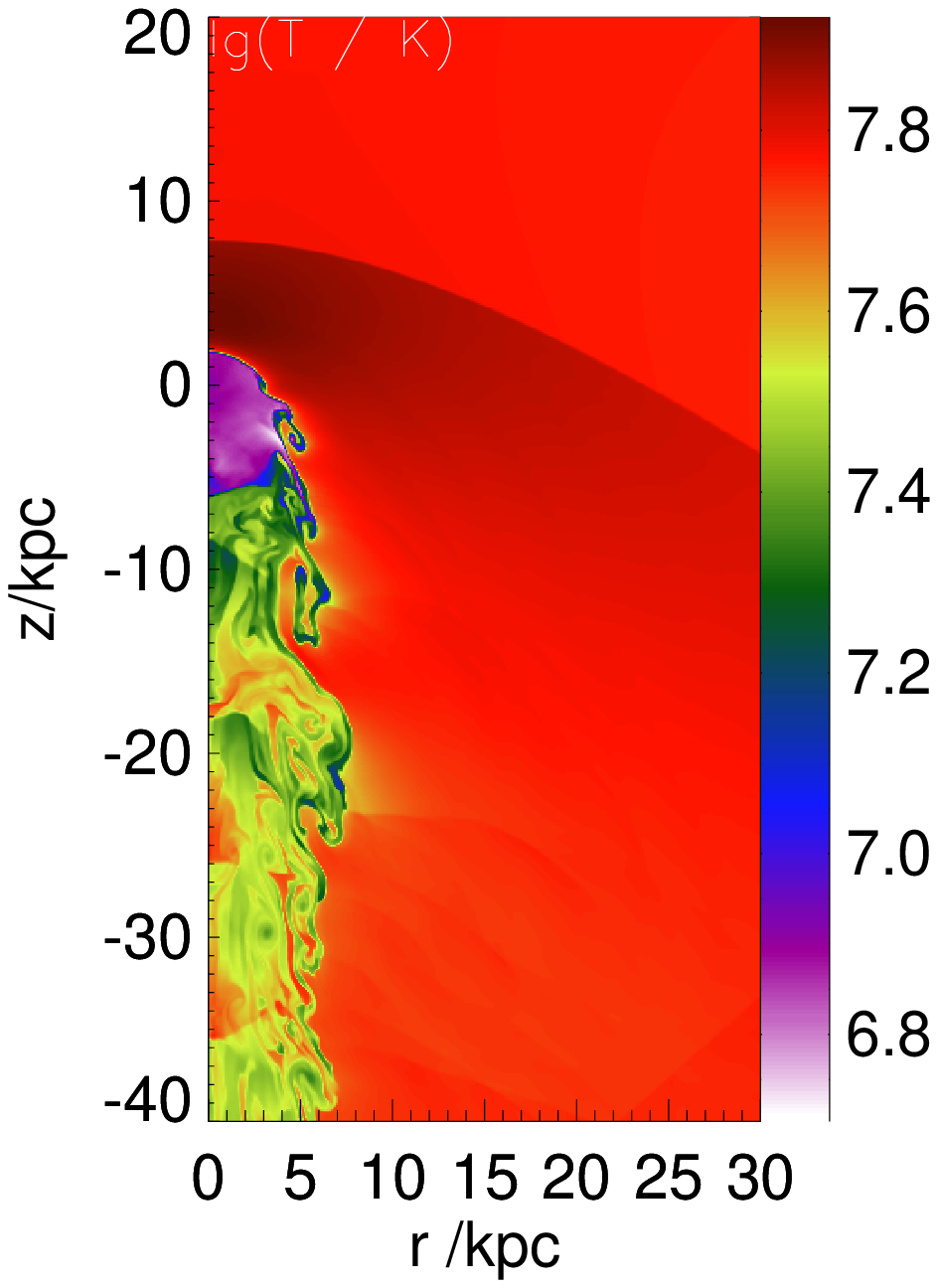}
   \end{minipage}
   \caption{Model $n_1T_2 \mathcal M_2 \beta_2$ seen in the Mach number,
            thermal pressure, density, and temperature.} 
\label{f:image_m}
\end{figure}

\begin{figure}
 \centering
 \includegraphics[width=0.8\linewidth]{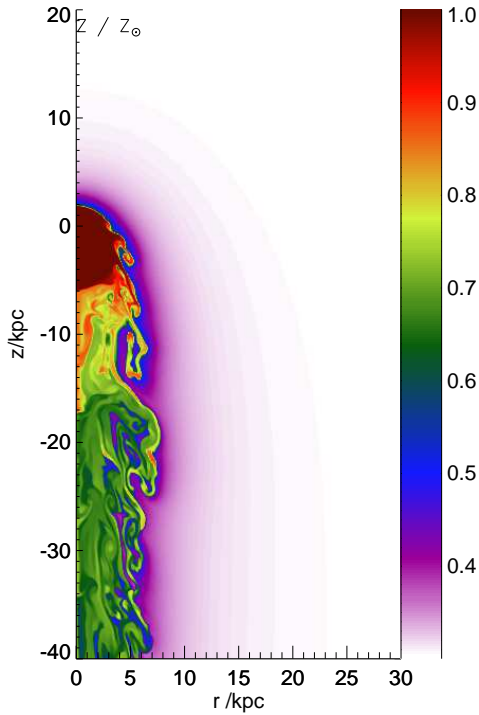}
 \caption{Iron abundance map for model $n_1T_2 \mathcal M_2 \beta_2$.}
\label{f:image_z}
\end{figure}

Fig.~\ref{f:image_m} shows a snapshot of the representative simulation case
$n_1T_2 \mathcal M_2 \beta_2$ in terms of the Mach number,
thermal pressure, density, and temperature distributions. 
At the very front of the corona is a smooth and distinct boundary that separates the corona from the ambient medium. 
This is a contact discontinuity, across which the density, temperature and metallicity change abruptly. 
Compared to the surrounding medium, the corona is cooler and denser. 
The iron abundance inside the corona is a constant which is equal to the value of injected material (Fig.~\ref{f:image_z}).  
Outside the corona the abundance drops rapidly to the value of the ICM, although it is contaminated by the local stellar feedback. 
Therefore, we can use the abundance to trace the morphology of the corona gas. 
Inside the corona, the Mach number of the gas flow is so low ($\sim0.1$) that it is almost hydrostatic (see \S~\ref{ss:interplay}). 

While the main body of a corona can reach a nearly steady state, both the interface
with the surrounding ICM and the tail are unstable. The individual features in 
these later parts can strongly fluctuate with time. The side horns are characteristic sign of the Kelvin-Helmholtz (KH) instability. 
As a result, the corona gas is torn off and pushed back to form a chaotic tail. 
Therefore, the stripping is primarily due to the hydrodynamic instability rather than the ram-pressure itself. 
Similarly, the instability also leads to the dynamic mixing of the corona gas with the ICM, although
numerically this is achieved on the spacial scale of the simulation resolution. 

\begin{figure}
 \centering
 \includegraphics[width=0.8\linewidth]{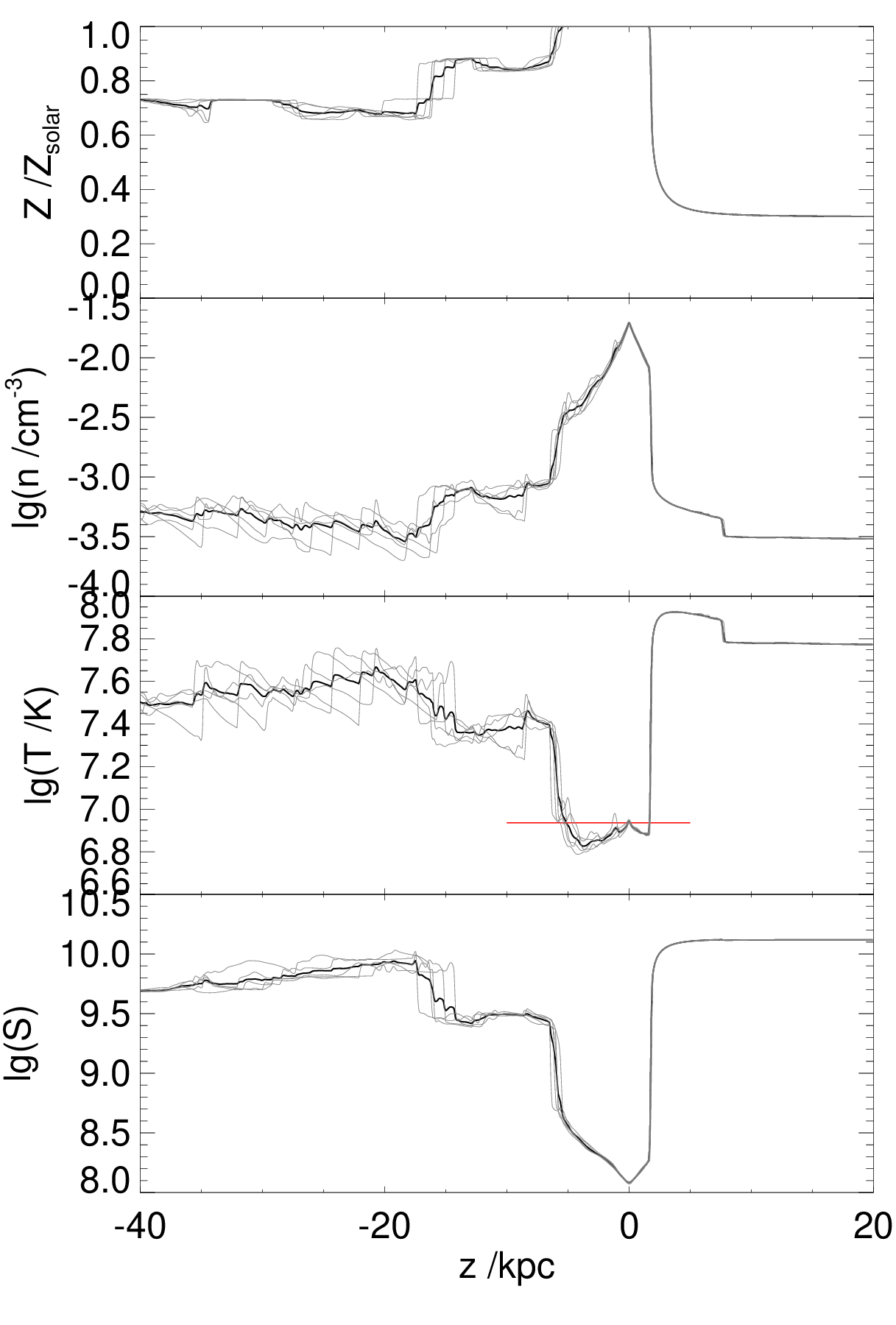}
 \caption{Iron abundance, denisty, temperature, and entropy distributions 
along z-axis of the simulation of case $n_1T_2 \mathcal M_2 \beta_2 $. 
The thick solid lines are time-averaged profiles over a time span from 250 to 300 Myr
(the residual wigglers are present due the the limited number of snapshots used in averaging),
while the thin lines are extracted from several snapshots. 
The horizontal red line in the temperature panel corresponds to $T={\beta\over 2.5k_B}$.} 
\label{f:cuts}
\end{figure}

Fig.~\ref{f:cuts} shows the 1-D distribution of the iron abundance, density, temperature and
entropy of the corona along the $z$ axis of the simulation box. Here the entropy is defined as $S = {T \over n^{\gamma-1}}$
($\gamma$ is the specific heat ratio of the gas and is equal to ${5\over 3}$).
The distributions are averaged over a time span of 50 Myr when the simulation has reached a quasi-steady state.

While the gas density drops substantially from the center of the spheroid to the 
outer outskirt of the corona, the temperature does not change much 
(Fig.~\ref{f:cuts}). The specific energy of the feedback determines the specific enthalpy 
of the corona gas and therefore the temperature ($T = {\beta \over 2.5k_B}$)
when the radiative cooling is not important as in the present case. The 
small drop of the temperature towards the outskirt (by a factor of up to 1.2)
is largely due to the outflow that needs to climb out of the gravity potential.
But, because of the distributed nature of the mass and energy injection, 
the drop is much smaller than what is predicted from the Bernoulli's law 
for ideal gas moving from the center to the outskirt. 

Another interesting characteristic of the simulated corona is the positive radial entropy gradient. 
This is apparently caused by the nearly constant temperature profile and 
the steep density drop from the center to the outskirts, as required to maintain a nearly hydrostatic state of the corona (\S~\ref{ss:interplay}).
Physically this positive entropy gradient is a natural result of an outflow that is continuously heated by the stars along the way out. 

\subsection{Similarity and Variance among the Cases}
\label{ss:model_v}

Here we focus on the similarity and significant variance in the hot gas 
properties of the various simulated cases, in reference to the 
representative one ($n_1T_2 \mathcal M_2 \beta_2$) detailed above.

\begin{figure}
 \centering
 \begin{minipage}[c]{0.48\linewidth}
  \centering
  \includegraphics[width=\linewidth]{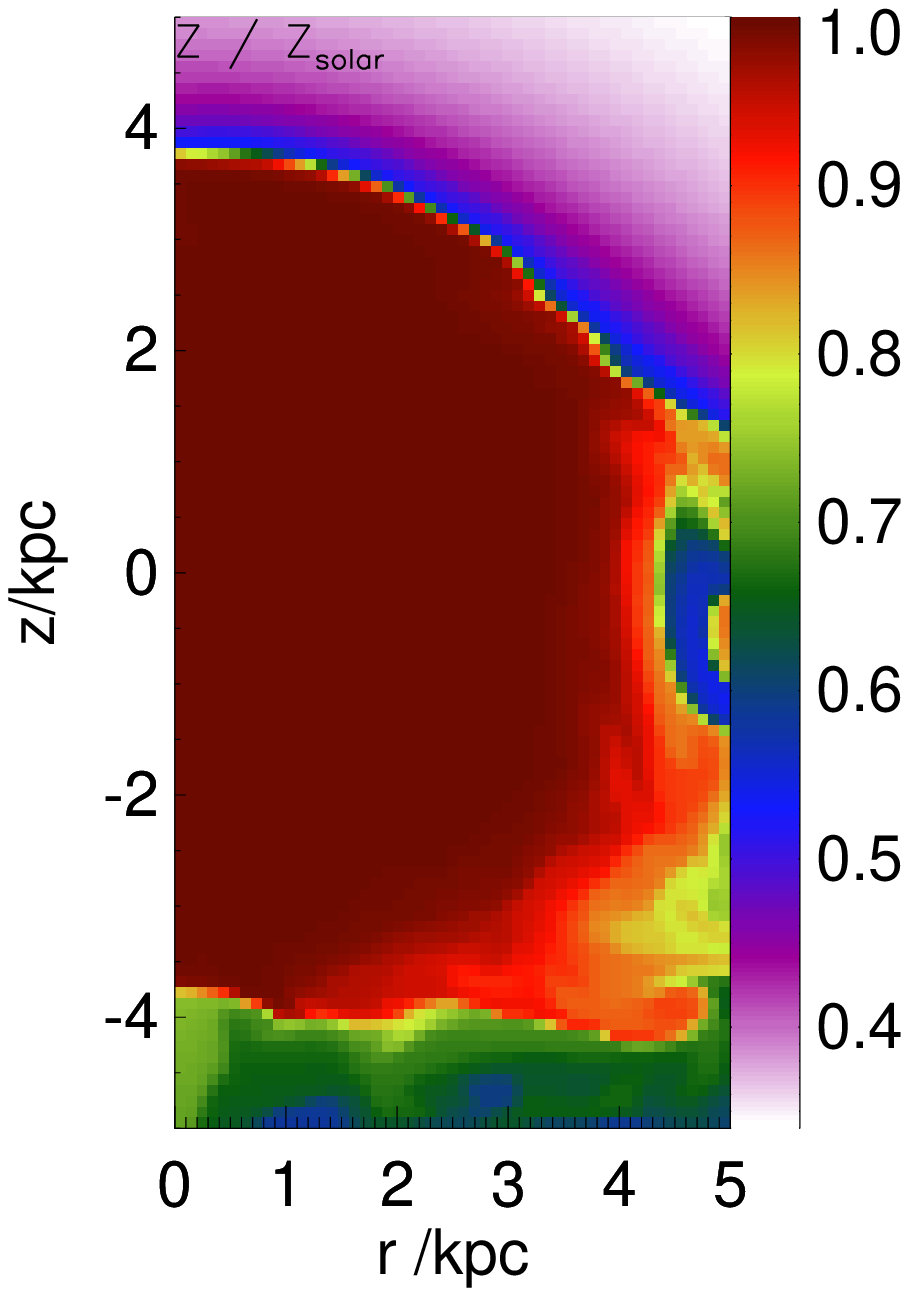}
 \end{minipage}
 \hspace{0cm}
 \begin{minipage}[c]{0.48\linewidth}
  \centering
  \includegraphics[width=\linewidth]{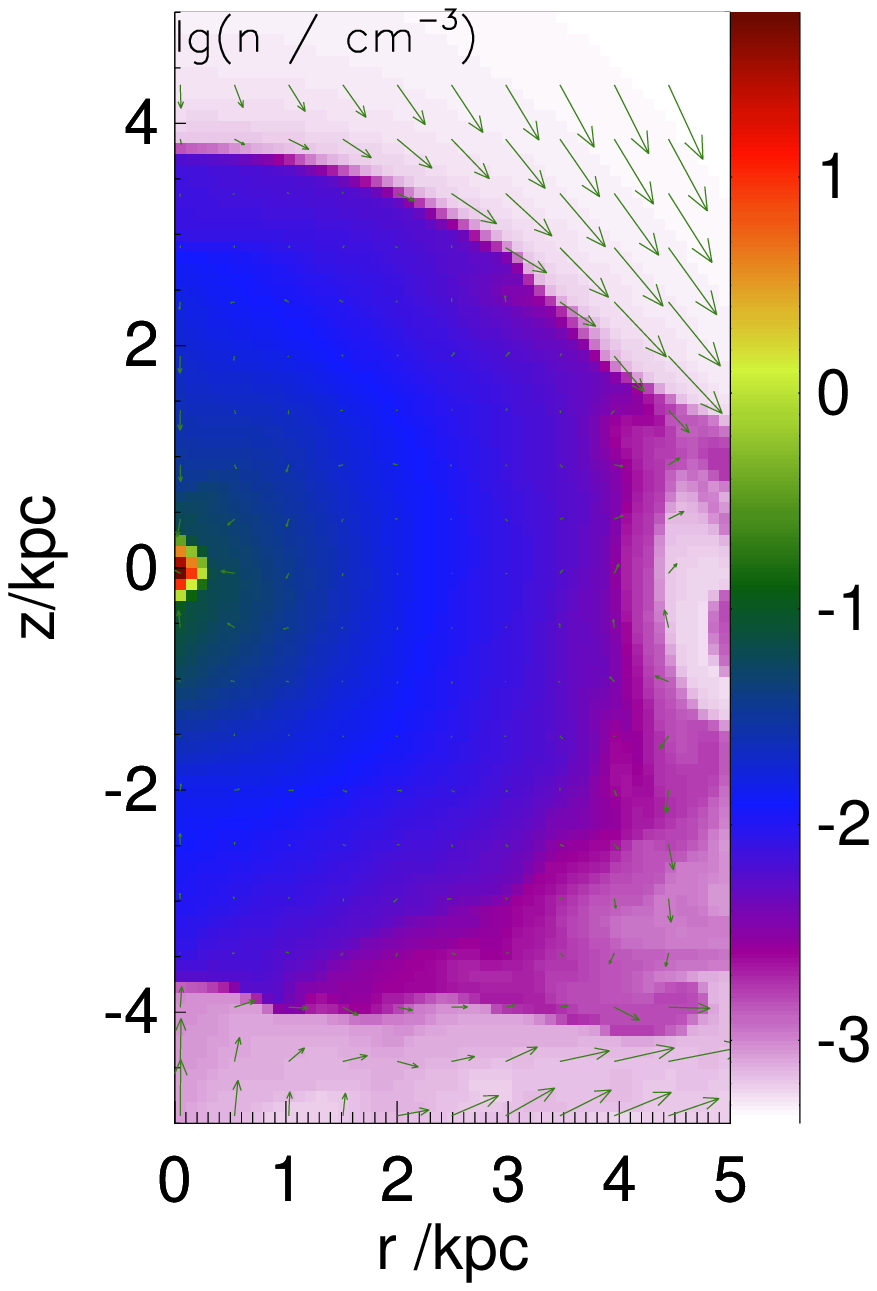}
 \end{minipage}
 \caption{Iron abundance and density distributions in the $n_{1}T_{2} \mathcal M_{1} \beta_{1}$ case.
          The arrows represent the velocity field. Note the central density peak (the red spot) at the very center.}
\label{f:image_cooling}
\end{figure}

In most of the simulation cases, the coronae are clearly in the outflow state. 
The radiative cooling is not important, except for some such as 
$ n_2T_1 \mathcal M_1 \beta_1 $, 
$ n_2T_1 \mathcal M_2 \beta_1 $,
$ n_1T_2 \mathcal M_1 \beta_1 $,
$ n_2T_2 \mathcal M_1 \beta_1 $ 
and 
$ n_2T_2 \mathcal M_2 \beta_1 $
, with combination of low specific energy and high ICM pressure. 
The gas at the spheroid center is so dense that an inward cooling flow is developed in the inner region.
(Fig.~\ref{f:image_cooling}). 
Such cooling flows, commonly seen in similar models and simulations, may 
naturally induce activities of the central supermassive black holes. 
The feedback from such activities has been proposed to substantially 
reduce the net cooling (e.g. \citealt{Mathews and Brighenti 2003, Fabian and Sanders 2009}). 
While a study of this topic is beyond the scope of the present work, 
we here keep our focus on discussing the morphological and physical properties of the outflow cases, 
which probably represent more typical cases of galactic coronae in the ICM. 

\begin{figure}
 \centering
 \begin{minipage}[c]{0.48\linewidth}
  \centering
  \includegraphics[width=\linewidth]{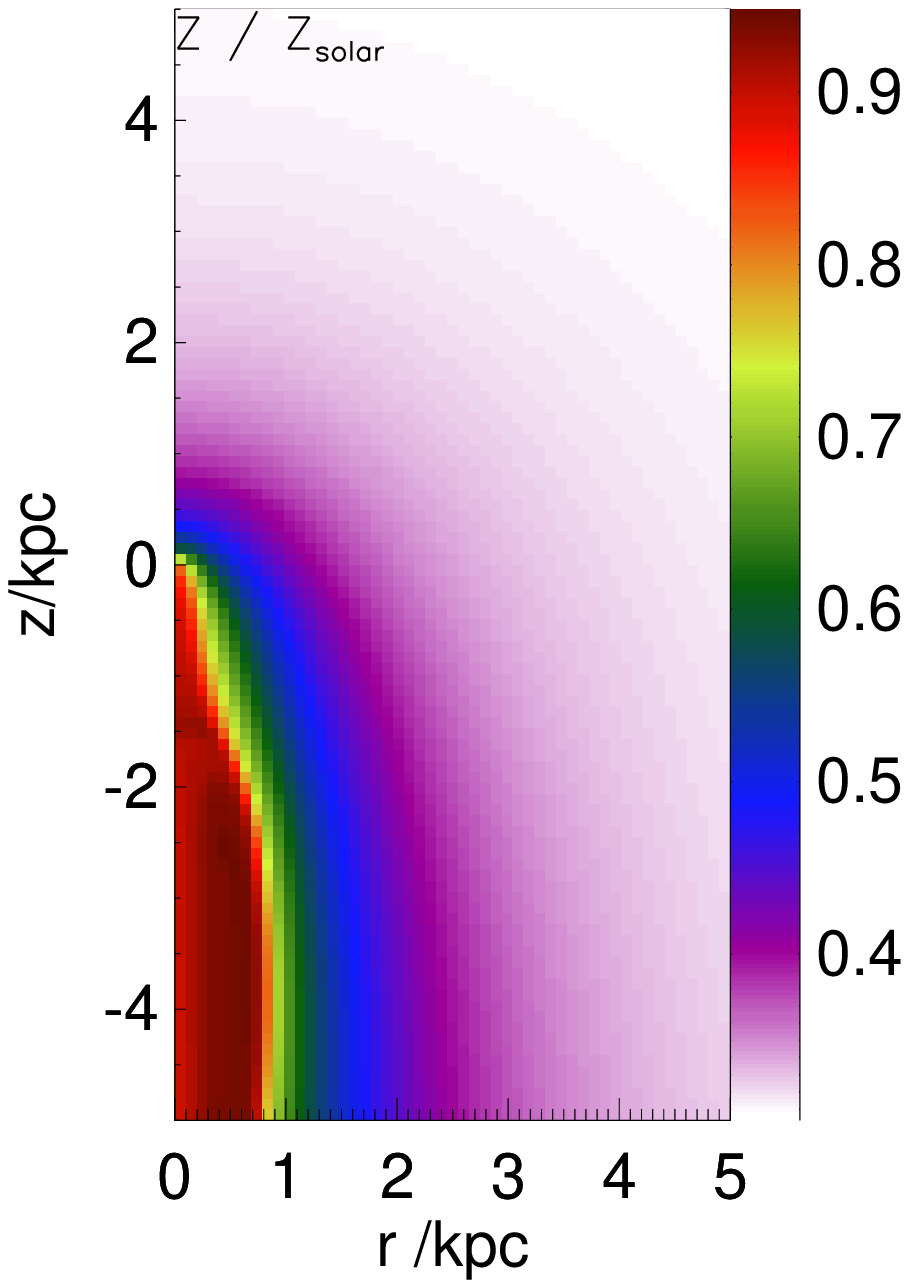}
 \end{minipage}
 \hspace{0cm}
 \begin{minipage}[c]{0.48\linewidth}
  \centering
  \includegraphics[width=\linewidth]{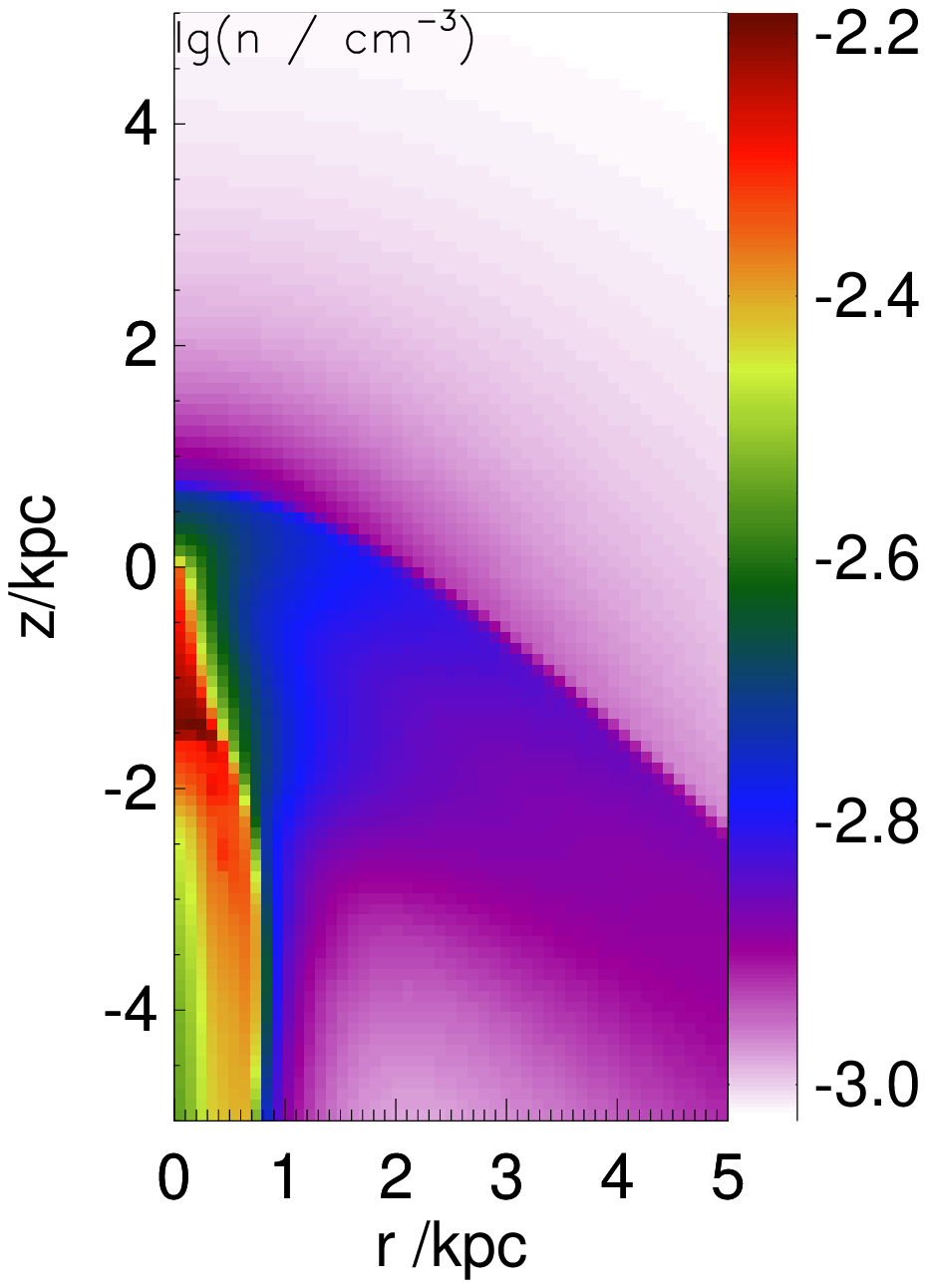}
 \end{minipage}
 \caption{Iron abundance and density distributions in the $n_{1}T_{2} \mathcal M_{1} \beta_{1}$ case, 
          in which the corona fails to form.}
\label{f:image_fail}
\end{figure}

For all the $n_2T_2 \mathcal M_3$ cases, coronae fail to form due to the high thermal and ram pressures, 
leaving only a track with high iron abundance (Fig.~\ref{f:image_fail}).  
In each of these cases, the density peaks away from the spheroid center.

\begin{figure}
   \centering
   \includegraphics[width=1.0\linewidth]{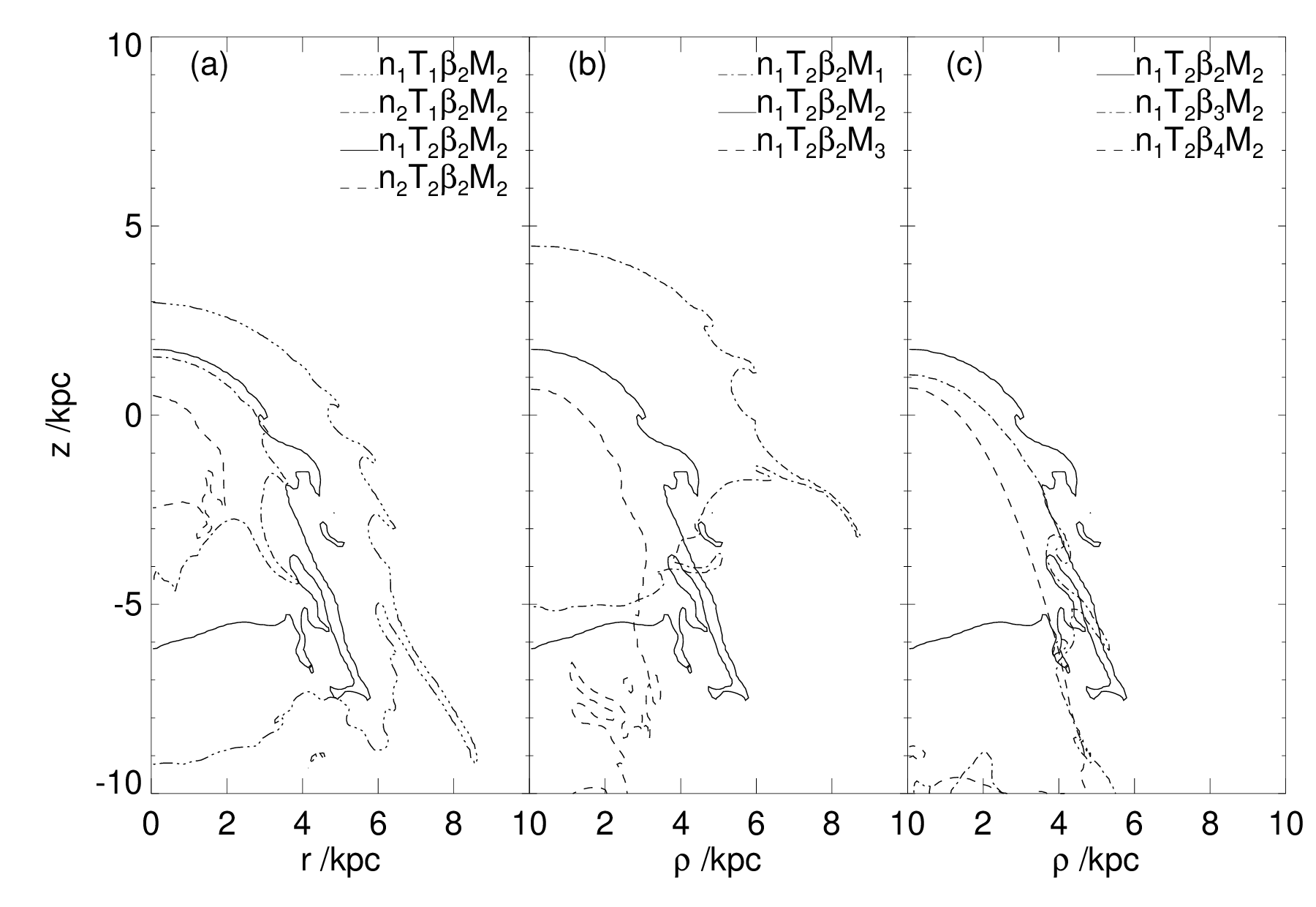}
   \caption{Outlines of the simulated coronae. The outline for each case is defined to be the contour of the iron abundance equal to $Z_{\odot}$. 
   The reference model $n_1T_2 \mathcal M_2 \beta_2$ is represented with solid line in all the three figures. 
   Panels (a), (b) and (c) correspond to variation in $nT$, Mach number and $\beta$, respectively, illustrating how the coronae respond to 
   the changes of the surrounding environment and stellar feedback.}
   \label{f:image_outline}
\end{figure}

Fig.~\ref{f:image_outline} show outlines of the coronae. 
These outlines are represented by the iso-abundance contours of the value of $Z_{\odot}$, 
or approximately the contact discontinuity between corona gas purely ejected by the spheriods and the ICM. 
The characteristic size of a corona is sensitive to the thermal pressure of the surrounding ICM 
as can be seen in panel (a) of Fig.~\ref{f:image_outline}. The corona in the ICM of $n_1T_1$, which is typical of 
a group or poor cluster, is more than $10$kpc across, while the corona in the ICM of $n_2T_2$, which is typical of 
the core region of a relatively rich cluster, is less than $3$kpc across. 
The motion of the host galaxy relative to the surrounding medium, which is represented by the Mach number, 
mainly influences lopsidedness of a corona. In the subsonic case, the corona is almost spherical,
while in the supersonic case with the Mach number as high as $1.8$, the corona is significantly narrowed 
and elongated.
As the input energy of the stellar feedback increases, the lopsidedness increases. 
This is not surprising. As the stellar feedback becomes more energetic, the density of the corona gas will be lower.
As a result, the gravitational restoring force become less important, compared with the ram pressure. 

\begin{figure}
   \centering
   \includegraphics[width=1.0\linewidth]{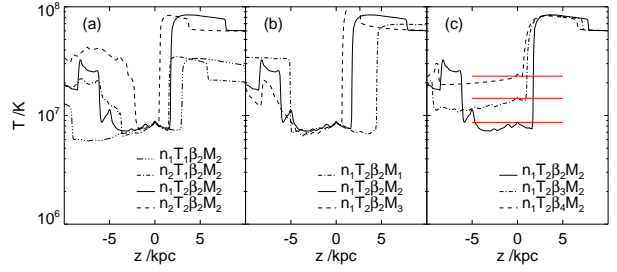}
   \caption{Temperature profiles along z-axis. The red horizontal lines in (c) represent ${\beta \over 2.5k_{B}}$.
   The reference model $n_1T_2 \mathcal M_2 \beta_2$ is represented with solid line in all the three figures. 
   Panels (a), (b) and (c) correspond to variation in $nT$, Mach number and $\beta$ respectively, illustrating how the coronae respond to 
   the changes of the surrounding environment and stellar feedback.}
   \label{f:plot_temp}
\end{figure}

The dependence of the corona temperature on the specific energy of injected material can be clearly seen in Fig.~\ref{f:plot_temp}. 
The peak value of the corona temperature is always roughly ${\beta \over 2.5k_{B}}$; it drops slightly outwards largely 
due to the presence of gravitational potential as described in \S~\ref{ss:model_r}.

\begin{figure}
   \centering
   \includegraphics[width=0.8\linewidth]{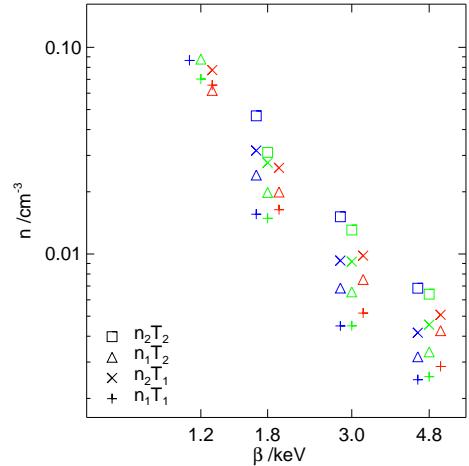}
   \caption{Peak density values of the coronae. 
   Cases with central cooling flow and those in which the coronae fail to form are not shown.
   Cases with different ICM thermal states are represented with different symbols, while different Mach numbers are coded in different colors
   with red for $1.8$, green for $1.2$ and blue for $0.6$. To avoid overlap among the symbols, the higher Mach number models 
   are shifted to the right a litte bit and the subsonic cases to the left.}
   \label{f:plot_dens_01}
\end{figure}

\begin{figure}
   \centering
   \includegraphics[width=1.0\linewidth]{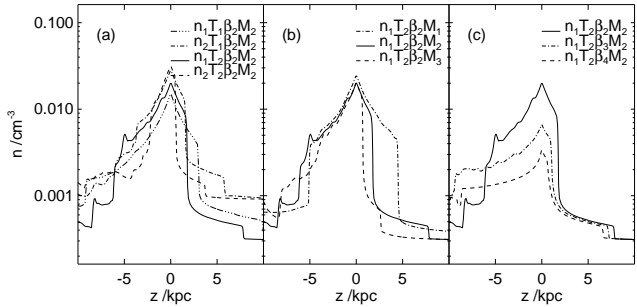}
   \caption{Density profiles along z axis. 
   The reference model $n_1T_2 \mathcal M_2 \beta_2$ is represented with the solid lines. Panels (a), (b) and (c) show the dependence on $nT$, Mach number, 
and $\beta$, respectively.}
   \label{f:plot_dens_02}
\end{figure}

We plot the peak density values of the coronae in Fig.~\ref{f:plot_dens_01} and selected density profile along the z-axis in Fig.~\ref{f:plot_dens_02}, 
illustrating how the coronae respond to
   the changes of surrounding environment and inner stellar feedback. 
Generally, the corona density depends strongly on the thermal pressure of the surrounding medium, but only weakly on the ram pressure.
In our simulation, the the Mach number ranges from $0.6$ to $1.8$, with the ratio between the largest ram pressure and the smallest one as high as $9$, 
the corona density changes slightly except for case $n_2T_2\mathcal M_1 \beta_2$ and its high Mach number version $n_2T_2\mathcal M_2 \beta_2$.
This makes sense because unlike thermal pressure, which compresses a corona from all directions, ram-pressure only acts on the front side and therefore mostly pushes the gas backwards rather than compressing it.
The coronae in cases of $n_2T_1$ are compressed more than those in cases of $n_1T_2$, although the ICM thermal pressure is the same.
This is caused by larger focusing effect of gravitational force of the galaxy on the ICM with lower temperature. 

\begin{figure}
   \centering
   \includegraphics[width=0.8\linewidth]{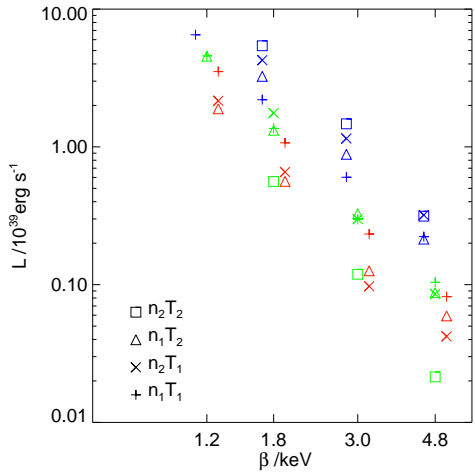}
   \caption{Luminosity of the coronae. 
   Cases with central cooling flow and those in which the coronae fail to form are not shown.
   Cases with different ICM thermal states are represented with different symbols, while different Mach numbers are coded in different colors
   with red for $1.8$, green for $1.2$ and blue for $0.6$. To avoid overlap among the symbols, the higher Mach number models 
   are shifted to the right a litte bit and the subsonic cases to the left.}
   \label{f:plot_lumin}
\end{figure}

Fig.~\ref{f:plot_lumin} includes the luminosity of each simulated corona in the $0.3-2.0{\rm~keV}$ 
X-ray band. For simplicity, we use the iron abundance distribution to define the shape of a corona
as what we do to plot the outline of a corona. Every grid point with iron abundance equal to the solar value
is included to measure the luminosity. 
The emissivity is a function that depends on both temperature and metallicity. 
If we assume the metals, mostly iron, produced by Ia SNe, are fully mixed in the corona gas, 
the total luminosity would be enhanced by a factor of about $3$.  
It is clear that the luminosity decreases with the increasing Mach number and/or $\beta$. 
But the dependence on the ICM thermal pressure is not that simple: 
the luminosity tends to increase with the pressure in subsonic cases (lower Mach numbers), 
while the trend goes in the opposite direction in the supersonic cases.


\section{Discussion}
\label{s:discussion}

The above results give a basic characterization of galactic coronae powered by
stellar feedback and semi-confined by the thermal and ram prssures of the ICM.
In this section, we first give a physical account of the apparent ICM impacts
on the corona characteristics as described above, then compare the results with 
observations to constrain the stellar feedback, and finally 
discuss the implications for other galactic properties.

\subsection{ICM Impacts on the Coronae}
\label{ss:interplay}
The properties of a corona are affected by several competing processes, 
the stellar feedback, the galaxy gravitational attraction, and the thermal/ram pressures of 
the ICM. 
In the simplest case, when the gravity and external pressure can be neglected, 
the density at the center would then be determined entirely by the mass/energy injection. 
In this case, we can define a characteristic density as
\begin{equation}
 \rho_{c,1} = {3 \over 4 \pi} {\dot M \over a^3} \tau.
\end{equation}
Here $\tau = {a \over c_s}$ is the dynamic time scale of the corona, where $c_s$ is the sound speed. 
Considering that the temperature is determined by the specific energy $\beta$, the characterisitic value of the pressure is 
\begin{equation}
 P_{c,1} = {3 \over 4 \pi} {\dot M \over a^2} {\sqrt{(\gamma-1)\beta} \over \gamma}.
\end{equation}

\begin{figure}
   \includegraphics[width=1.0\linewidth]{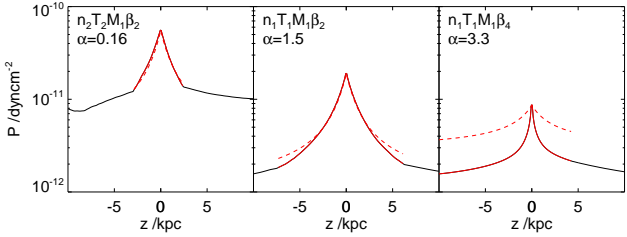}
   \caption{Comparison between numerical pressure profiles and the theoretical profiles with isothermal assumption. 
   The solid lines are the pressure profiles extracted from the simulations and the dashed lines represent the corresponding hydrostatic 
   solutions. The region inside the coronae is colored in red. In each of the panels, the peaks of both of the numerical 
   profile and theoretical one are positioned at the same point. }
   \label{f:plot_compare}
\end{figure}

In the other extreme, when the gravity is important (i.e.,
the corona is nearly hydrostatic), the Mach number of the outflow must be low. 
Clearly, in this case, the external thermal pressure becomes important as well.
Because the corona is nearly isothermal, the pressure distribution is
\begin{equation}
 \ln(P) = \ln(P_0) - {\mu m_p (\Phi-\Phi_0) \over k_B T},
\end{equation}
where $P_0$ is the pressure at the outer boundary and $\Phi-\Phi_0$ is the 
gravitational potential difference.
Assuming that the corona size is considerably larger than the 
scale $a$ of the stellar spheroid, we obtain a characteristic peak 
pressure as
\begin{equation}
 P_{c,2} = P_{ICM} \exp \Biggl({ {GM_s\over a} \sqrt{ {\gamma \over (\gamma-1)\beta} } }\Biggr).
\label{e:hs}
\end{equation}
To characterize the relative importance of the feedback to the gravity, 
we define a dimensionless parameter as 
\begin{equation}
 \alpha = {P_{c,1} \over P_{c,2}}.
\end{equation}
The values of this parameter for the simulated cases are 
listed in Table.2.
\begin{table}
\label{t:alpha}
 \centering
 \begin{minipage}[c]{0.8\linewidth}
  \centering
  \caption{$\alpha$ for various cases.}
  \begin{tabular}{@{}lcccc@{}}
 \hline
 \hline
            & $\beta_1$  & $\beta_2$  & $\beta_3$  & $\beta_4$  \\
 \hline
 $n_1,T_1$  & $0.99$     & $1.47$     & $2.30$     & $3.33$     \\
 $n_2,T_1$  & $0.33$     & $0.49$     & $0.77$     & $1.11$     \\
 $n_1,T_2$  & $0.33$     & $0.49$     & $0.77$     & $1.11$     \\
 $n_2,T_2$  & $0.11$     & $0.16$     & $0.25$     & $0.37$     \\
 \hline
 \hline 
\end{tabular}
\end{minipage}
\end{table}
Fig.~\ref{f:plot_compare} compares the hydrostatic solutions (Eq.~\ref{e:hs})  
to the simulations with three different $\alpha$ parameters. 
In those cases with small $\alpha$, such as $n_2T_2\mathcal M_1 \beta_2$, in which the gravity dominates over the feedback, the hydrostatic solutions give nearly perfect matchs to the simulated pressure profiles. While in a case like 
$n_1T_1\mathcal M_1 \beta_4$ (with $\alpha=3.3$), where the feedback dominates, the deviation of the hydrostatic
solution from the simulated profile is apparent.
For the cases which we think are plausible ($\beta$=1.8keV, see \S~\ref{ss:discussion_02}), 
$\alpha$ is about $1.0$ or much less than $1.0$, and the coronae are largely in hydrostatic state.

Although the same model galaxy (in terms of the stellar and dark matter masses)
is adopted for all the simulation cases, the resultant X-ray luminosities 
can still differ by up to two orders of magnitude, due to the different 
choices of the Mach number, $\beta$, and/or thermal pressure values of the ICM.
Because the corona temperature is determined by $\beta$, the gas density is 
\begin{equation}
n_{c} \sim {P_{ICM} \over \beta},
\end{equation}
if the corona is in a nearly hydrostatic state.
Therefore, the surface brightness of a corona provides a measure of the 
ambient ICM pressure and thus
may be used to estimate the line-of-sight position in a cluster.

However, the X-ray luminosity of a corona depends on several factors.
Fig.~\ref{f:plot_lumin} shows a clear anti-correlation between the luminosity 
and specific energy $\beta$. This anti-correlation is primarily due to the density
decrease with the increase of $\beta$, although it does not strongly 
affect the size of a corona. The ICM thermal pressure tends to squeeze the corona, 
hence enhance its luminosity (subsonic cases). But this effect is complicated by the presence of the
ram-pressure. 
As the Mach number increases, 
the lumonosity can decrease because the ram-pressure stripping reduces the overall size ofthe corona. 
These dependences on the environment as well as 
the stellar feedback energetics may naturally explain the observed large dispersion of $L_x/L_K$ 
for spheroids of similar $L_K$. The complications in the dependences may also account for 
the lack of a clear observed trend in
the ICM environment effect on X-ray luminosities of coronae 
\citep{Sun et al. 2007, Mulchaey and Jeltema 2010}.

\subsection{Implication for the Feedback Model}
\label{ss:discussion_02}

In our model of the galactic coronae, the gas temperature is primarily determined by the
specific energy of the stellar feedback and thus should not change significantly with the stellar mass. 
This independence on the mass or $L_K$ is consistent with 
the temperature measurements of the coronae of intermediate-mass spheroids 
\citep{David06, Sun et al. 2007, Jeltema et al. 2008, Boroson et al. 2010}. 
This is in contrast to the correlation between the 
temperature and $L_{B}$ for more massive systems such as clusters and groups of galaxies \citep{Helsdon and Ponman 2003}.  
In particular, galaxy clusters show a well-defined 
scaling law between the temperature and luminosity of 
the observed hot gas, which is a natural result of the predominant gravitational heating 
in the self-similar cluster formation. The scaling law for lower mass systems (e.g., groups of galaxies)
is known to be slightly different from that for clusters (e.g., showing an 'entropy floor'), 
which is believed to be an imprint of preheating(e.g., starburst and early AGNs).
Correlation between $L_{B}$ and the temperature of hot gas is observed in massive X-ray-bright elliptical galaxies, 
especially for central galaxies in groups and clusters (e.g,, \citealt{O'Sullivan et al. 2003}).
But the entropies are found lie below the entropy floor($\sim 10^{9}\rm~K~cm^{-2}$) discovered in groups of galaxies.
Radiative cooling could account for the low entropy, 
although how this runaway process may be balanced 
by the heating due to the mechanical inputs from both stellar and AGN feedbacks remains unclear. 
We have shown that the low value and the positive radial gradient of the entropy 
are expected from the distributed feedback in intermediate-mass spheroids, 
in which the radiative cooling is not important.
Therefore, we may conclude that the coronae of intermediate-mass spheroids represent the extreme case in which 
the stellar feedback plays a dominant role, 
which means they are produced by stellar mass loss and heated by SNe.

We may further constrain the stellar feedback based on the 
measured temperature of the coronae.
Though with a relative large dispersion, the measured temperatures
are mostly fall in the range of $0.5 - 1.0{\rm~keV}$, which 
is significantly higher than those measured in
field spheroids, but is still substantially lower than what is 
inferred from our 2-D simulations if the canonical specific energy 
value of the stellar feedback is assumed ($\sim 5{\rm~keV}$; see \S~\ref{s:setup}). 
Part of this discrepancy could still be due to the 3-D effects of discrete
heating by Ia SNe, as mentioned in \S~1 (and characterized in \citealt{Tang09b,TangWang10}).
But we expect that such effect should be substantially weaker in the compact 
coronae embedded in the high pressure ICM and that the measured 
temperature should more faithfully reflect the specific energy of the feedback. 
To match the measured temperature range of the coronae requires a specfic energy 
of $\sim 1.5 - 3 {\rm~keV}$, or a factor of $\sim 2-3$ lower than the canonical value 
($5{\rm~keV}$; \S~\ref{s:setup}). This factor is probably still within the uncertainties
of the semi-emipirical mass and energy injection rates. Further, the 
assumed mechanical energy per Ia SN could be somewhat (e.g., a factor of $\sim 2$) 
less than $10^{51} {\rm~ergs}$. Also a considerable fraction of the energy 
can be used to generate cosmic rays, magnetic field and turbulent
motion. The diversion of the energy into these various forms could significantly 
reduces the temperature, although 
the hydrodynamics of the coronae, hence the density and pressure distributions,
should not be significantly affected.
The simulated coronae with $\beta = 1.8{\rm~keV}$ generally
have individual luminosties of a few times $10^{39}{\rm~erg~s^{-1}}$, 
consistent with the observed range of 
$10^{39}\sim10^{40}{\rm~erg~s^{-1}}$ \citep{Sun et al. 2007}. A considerably large
value of $\beta$ is not favored, because it decreases the expected luminosity steeply
(Fig.~\ref{f:plot_lumin}).

\subsection{Implifications for Understanding Other Galactic Components}
We discuss here the potential impacts of the pressure or density enhancement 
of the coronae on the fueling of the central SMBHs and the evolution 
of cool gas, if present in the spheroids. 

\begin{figure}
   \centering
   \includegraphics[width=0.8\linewidth]{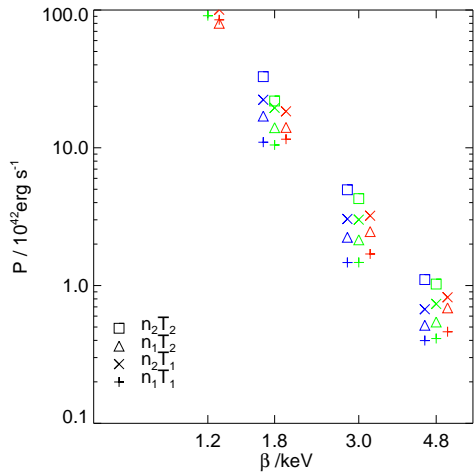}
   \caption{Estimated AGN power. 
   Cases with cooling flow and those in which the coronae fail to form are not shown.
   Different ICM thermal states are represented with different symbols and different Mach numbers are coded in different colors
   with red for $1.8$, green for $1.2$ and blue for $0.6$. To avoid overlap among the symbols, the higher Mach number models 
   are shifted to the right a litte bit and the subsonic cases to the left.}
   \label{f:plot_agnpower}
\end{figure}

The simulation shows that the central density of a corona is sensitive to the thermal pressure of the surrounding medium. 
To infer the power of a SMBH, we adopt the Bondi accretion rate \citep{Edgar 2004}:
\begin{equation}
\dot M = {4\pi G^2 M_{BH}^2 \rho \over c_{s}^3}.
\end{equation}
where $M_{BH}$ is the mass of the SMBH, while  $\rho$ and $c_{s}$ are the density and sound speed at the center of a corona. 
Assuming the fraction of the accretion energy released is $\eta=0.1$, the power of the SMBH can be approximated as
\begin{equation}
2.12 \times 10^{41}{\rm~erg~s^{-1}} \Bigl({M_{BH}\over 10^8{\rm~M_{\odot}}}\Bigr)^2 \Bigl({n\over {\rm~cm^{-3}}}\Bigr) \Bigl({T \over 7.0\times 10^7{\rm~K}}\Bigr)^{-1.5}
\end{equation}
The SMBH mass can be estimated from its correlation with the spheroid mass $M_{BH}\sim0.006M_{bulge}$ \citep{Magorrian et al. 1998}.
Fig.~\ref{f:plot_agnpower} shows the dependence of the power on the ICM state, the Mach number of spheroid, and the specific energy of the feedback. 

In particular the luminosity for the most plausible $\beta_2$ cases is about $10^{43} {\rm~erg~s^{-1}}$, 
which falls in the range of the X-ray power of a low luminosity
AGN. This indicates that a compact corona built up by stellar feedback
and embedded in cluster environment could feed a moderate AGN. The ICM
pressure also tends to enhance the accretion, consistent with the
finding that the galaxies with $L_x >
10^{42}{\rm~erg~s^{-1}}$ AGNs are more centrally concentrated than
ones without \citep{Martini et al. 2007}.
These galaxies with AGNs are not dominated by galaxies that have
recently entered the clusters. 
Similar conclusions are also reached in more recent studies, such as
the one by \citet{Hart et al. 2009}, based on the analysis
of a sample of $P_{1.4GHz}>3\times10^{23}{\rm~W Hz^{-1}}$ radio galaxies and $L_{0.3-8keV}>10^{42}{\rm~erg~s^{-1}}$ point sources. 

The pressure enhancement could also have significant impacts on cool
gas in a galaxy. Under high pressure, cool gas exists
preferentially in molecular form rather than atomic one. The
compression of cold gas because of the ICM pressure could further 
lead to star formation \citep{Bekki and Couch 2003},  depriving the
galaxy of the gas further. 
Thus it is expected that galaxies contain less amounts of cool gas 
in clusters than in the field. These impacts should affect cool gas
not only in spheroids, but in spirals as well, consistent with
existing observations \citep{Young and Scoville 1991}. 


\section{Summary}

We have conducted a range of 2-D hydrodynamic simualtions of galactic
coronae that result from gradual energy and mass feedback in 
stellar spheroids moving in the ICM enviornment. We have focused on spheroids that
are in the intermediate-mass range (corresponding to 
$L_K \sim 10^{11} - 10^{12}L_{K,\odot}$) so that both the AGN feedback and the
radiative cooling of the hot gas could largely be neglected.
We explore the dependence of corona properties on the specific energy of the stellar feedback 
as well as on the ram and thermal pressures of the ICM.
Our major results and conclusions are as follows:

\begin{enumerate}
\item X-ray coronae embedded in clusters could be naturally explained by the
 subsonic outflow driven by the stellar feedback, semi-confined by
 the ram-pressure and compressed by thermal pressure of the surrounding ICM. 
The corona temperature depends primarily on the specific energy of the
input material in such a way that $T \sim {\beta \over 2.5k_B}$. The
decrease of the thermal energy due to the climbing of the
gravitational potential and the expansion is largely compensated by
the distributed heating by Ia SNe. This result naturally explains the
lack of the correlation between  the temperature and K-band luminosity for
the spheroids in our considered mass range. 
An outflow powered by a distributed feedback also has a positive radial entropy profile, 
mimicking what may be produced by a "cooling flow". 

\item The coronal gas is typically in an approximate hydrostatic
  state. As a result, the density of the corona gas 
depends strongly on the thermal pressure of the ICM, but only weakly
on the ram pressure. Therefore, the surface brightness of X-ray
emission is a good measurement of the thermal ICM pressure, which may
be used to estimate the line-of-sight location of a spheroid in a
cluster. The total X-ray luminosity of a corona decreases with the increase of the feedback
specific energy. 
The thermal pressure tends to increase (or reduce) the luminosity in subsonic (supersonic) cases.

\item The semi-confinement of the coronae by the ICM allows a good
constraint on the energetics of the stellar feedback. To be
consistent with the observed X-ray luminosity and 
temperature, the specific energy  of the feedback should be $\sim
1.5-3{\rm~keV}$, a factor of 2-3 smaller than the value inferred from the
commonly accepted semi-empirical Ia SN and mass-loss rates, assuming
the mechanical energy of $10^{51} {\rm~ergs~s^{-1}}$ per SN.

\item The relatively high pressure of the coronae in the ICM may have important implications 
for understanding the AGN activity as well as the cool gas properties in spheroids. 
The density increase caused by the ICM pressure, for example, could
enhance the Bondi accretion, which may explain the observed central
concentration of  AGNs in clusters.  The high pressure can further
compress the gas to form molecular clouds and enhance star formation.
The combination of the enhanced consumption and the
ram-pressure striping can naturally lead to the deprivation of gas and
subsequent passive evolution of galaxies in clusters.
  
\end{enumerate}

\section{Acknowledgements}
We thank S.-K. Tang for his help in the initial setting up of the
simulations. The software used in this work was in part developed
by the DOE-supported ASC/Alliance Center for Astrophysical
Thermonuclear Flashes at the University of Chicago.
Simulations were performed at the Pittsburgh Supercomputing Center
supported by the NSF. The project is partly supported by NASA through grant NNX10AE85G.


\end{document}